\documentclass[aps,prd,10pt,twocolumn]{revtex4}

\usepackage{epsfig}
\usepackage{bm}
\usepackage{latexsym}
\usepackage{natbib}
\usepackage{url}
\usepackage{dcolumn}
\usepackage{color}
\usepackage{amsfonts,amssymb,amsmath}
\usepackage{graphicx,epsfig}
\usepackage{psfrag}
\usepackage{subfigure}
\usepackage{hyperref}
\hypersetup{colorlinks=true}

\newcommand{\ba}{\begin{array}}
\newcommand{\ea}{\end{array}}
\newcommand{\be}{\begin{equation}}
\newcommand{\ee}{\end{equation}}
\newcommand{\bea}{\begin{eqnarray}}
\newcommand{\eea}{\end{eqnarray}}

\begin{document}

\title{Constraints on cosmological viscosity and self interacting  dark matter from gravitational wave observations}

\author{Gaurav Goswami$^{a,}$\footnote{gaurav.goswami@ahduni.edu.in}}
\author{Girish Kumar Chakravarty$^{b,}$\footnote{girish20@prl.res.in}}
\author{Subhendra Mohanty$^{b,}$\footnote{mohanty@prl.res.in}}
\author{A. R. Prasanna$^{b,}$\footnote{prasanna@prl.res.in}}
\affiliation{$^a$ School of Engineering and Applied Science, Ahmedabad University, Ahmedabad 380009, India \\
$^b$ Theoretical Physics Division, Physical Research Laboratory, Ahmedabad 380009, India }

% \title{Constraints on cosmological viscosity from GW150914 observation}
% 
% \author{Gaurav Goswami$^{a}$, Girish Kumar Chakravarty$^{b}$, Subhendra Mohanty$^{b}$ and A. R. Prasanna$^{b}$}
% 
% \affiliation{$^a$ School of Engineering and Applied Science, Ahmedabad University, Ahmedabad 380009, India \\
% $^b$ Theoretical Physics Division, Physical Research Laboratory, Ahmedabad 380009, India}
% \\
% $^3$ Dipartimento di Fisica "E.R. Caianiello" Universit\'a di Salerno, I-84084 Fisciano (Sa), Italy \\
% $^4$ INFN - Gruppo Collegato di Salerno, Italy}

% \vspace*{1cm}
% \title{Constraints on cosmological viscosity from GW150914 observation}
% \bigskip
% \author{Gaurav Goswami}
% \email{gaurav.goswami@ahduni.edu.in}
% \affiliation{School of Engineering and Applied Science, Ahmedabad University, Ahmedabad 380 009, India}
% \author{Subhendra Mohanty}
% \email{mohanty@prl.res.in}
% %\affiliation{Physical Research Laboratory, Navarangpura, Ahmedabad 380009,India.}
% \author{A. R. Prasanna}
% \email{prasanna@prl.res.in}
% \author{Girish Kumar Chakravarty}
% \email{girish20@prl.res.in}
% \affiliation{Theoretical Physics Division, Physical Research Laboratory, Ahmedabad 380009, India}

\begin{abstract}
It has been shown that gravitational waves propagate through ideal fluids without experiencing any dispersion or dissipation. However, if the medium has a non-zero shear viscosity $\eta$ , gravitational waves will be dissipated at a rate proportional to $G \,\eta$. 
We constrain dark matter and dark energy models with non-zero shear viscosity by calculating the dissipation of gravitational waves from GW150914 which propagate over a distance of $410\, $ Mpc through the dissipative fluid and comparing the data with the theoretical prediction. 
This provides a proof-of-principle demonstration of the fact that future observations gravitational waves at LIGO have the potential of better constraining the viscosity of dark matter and dark energy.
%{\color{red} We put an upper bound on the shear viscosity of the cosmological fluid as 
%$\eta < 2.3 \times 10^{9} ~{\rm Pa \,\, sec}$ which is close to the critical viscosity of fluids at which the viscous pressure becomes significant for the 
%dynamics of the universe. Future observations of gravitational waves at LIGO have the potential of detecting  the viscosity of dark matter and dark energy.}  
\end{abstract}

\maketitle

%\section{Introduction}
%{\it Introduction:} 
The discovery of Gravitational Waves (GW) by the LIGO collaboration  \cite{LIGO} opens a new window for astronomy and cosmology. 
The source of GW150914 could also be the source of the 1 sec x-ray burst observed by Fermi GBM \cite{Connaughton:2016umz}
with a 0.4 sec delay w.r.t the GW event and with sky localization consistent with the LIGO observation. 
These measurements of gravitational waves and their possible electromagnetic counterparts  can tell us about the nature of the astrophysical sources \cite{source1,source2,source3,source4,source5,source6,source7,source8,source9,Belczynski:2016obo,Liu:2016olx},  
test general relativity \cite{gr1,gr2,gr3,gr4,gr5,gr6} and local Lorentz invariance \cite{LI}. 
Future observations of stochastic gravitational waves can tell us about the energy scales of the first order phase transitions in the early universe \cite{phase1,phase2}.   In this paper we study the effect of the medium on the propagation of gravitational waves with the aim of deducing the properties of dark matter and dark energy by studying the observed waveforms. It was shown by Ehlers et. al  \cite{Prasanna1, Prasanna2} in full generality that gravitational waves propagating through ideal fluids do not suffer any dispersion or dissipation. 
Prasanna  \cite{Prasanna3} generalized this treatment to the case of non-ideal fluids and showed that only the coefficient of shear viscosity affects the gravitational waves as they can be attenuated by the medium. 
This general conclusion agrees with the earlier derivations of attenuation of gravitational waves due to non-ideal fluid in a 
Friedman-Robertson-Walker (FRW) background 
\cite{Hawking, Dyson, Esposito, Weinberg, Madore, Anile} where it was shown that the attenuation length is $(16 \pi G \eta)^{-1}$ 
(in this context, see also \cite{Kocsis:2008aa}). 
Shear and bulk viscosity have been invoked to avoid initial singularity at the Big Bang \cite{sing1,sing2,sing3}, and as dark energy 
\cite{DE1,DE2,DE3}. Dark matter with self interaction i.e non-zero shear and bulk viscosity has been used  \cite{DM1,DM2,Fan:2013yva,Foot:2014uba,Foot:2016wvj} for explaining the lack of density spikes in the cores  \cite{Core-Cusp} or substructures \cite{Substructure}, or the paucity of dwarf satellite galaxies \cite{Missing-Satellites} which are seen in simulations with collision-less ideal fluid dark matter.  

In this paper we study this effect in the context of the recent observations of Gravitational Waves.
We consider the possibility that the analysis of the GW150914 could allow us to put observational constraint on the shear viscosity. We find that it is in-principle possible to constrain the shear viscosity of the cosmic fluid using GW observations and that the corresponding viscosity values lie in an interesting range which may be relevant to the dissipative dark matter and dark energy models.
It turns out that the dissipative dark matter in galaxy clusters such as Abell 3827 \cite{Massey:2015dkw} has the shear viscosity in the range constrained from the GW150914 analysis. 
Thus, in future, gravitational waves could possibly provide a good observable handle for the measurement of the viscosity of cosmological fluids.
We begin by deriving the wave equation for gravitational waves in a FRW universe filled with a viscous fluid.

 \vspace{.3cm}

{\it Gravitational wave propagation through a viscous fluid: }
The energy momentum tensor of a non-ideal fluid can be written in the general form
\be
T_{\mu \nu} \equiv (\rho+p) u_\mu u_\nu + p g_{\mu \nu} - 2 \eta \sigma_{\mu \nu} -\xi \theta \Delta_{\mu \nu}
\ee
where $\eta$ is the coefficient of shear-viscosity , $\xi$ is the coefficient of bulk viscosity , $\sigma_{\mu \nu}$ is the shear, $\theta$ is the volume expansion of the fluid and $\Delta_{\mu \nu}= g_{\mu \nu} + u_{\mu} u_\nu$ is the projection operator to project to subspace orthogonal to 
the fluid four velocity $u_\mu$. 

 Observations of the cosmic microwave background anisotropy \cite{Planck} shows that the universe can be described by a perturbed  Friedmann-Robertson-Walker (FRW) metric. In a FRW universe  with a non-ideal fluid,
the isotropy of the background ensures that the scalar and tensor perturbations evolve independently at linear order in perturbation theory \cite{Mukhanov}.
%Thus, neither the background dynamics nor the scalar perturbations (at linear order) can probe the effects of shear viscosity. 
The tensor perturbations at linear order can however probe the shear viscosity as we now show.
We thus consider the background FRW metric with only tensor perturbations
 \be
 ds^2= -dt^{2} + a^2(t) \left[\delta_{ij} + h_{ij}\right] dx^i dx^j \; ,
 \label{RW}
 \ee
where the tensor perturbations are transverse and traceless i.e. $\partial^i  h_{ij} = h^i_i=0$. 
We work with the units  $\hbar = c = 1$.
%and $M_{Pl}^2 = (8 \pi G)^{-1}$. 
 
 The total four velocity is $u_\mu = u^{(0)}_\mu + \delta u_\mu$, normalizing the four velocity i.e., 
 $g^{\mu \nu} u_\mu u_\nu=-1$, and retaining the first order terms in the metric and velocity perturbations gives us the relation 
 \be
 \delta u^{\mu}=-\frac{1}{2}g^{\mu\nu}\delta g_{\lambda\nu} \, u^{\lambda}
 \ee
 In the local rest frame of the fluid where $u^{\mu}\equiv\delta^{\mu}_{0}=(1,\vec 0)$, for the perturbed FRW metric 
 Eq (\ref{RW}), the velocity perturbations $\delta u^{\mu}$ vanish. Also, for the perturbed FRW metric (\ref{RW}) and in the local rest frame of the fluid, the bulk expansion rate $\theta=\nabla_{\mu} u^{\mu}$ and $ij-$component of the
 shear viscosity $\sigma_{\mu\nu} \equiv \frac{1}{2} \left[u_{(\mu;\nu)}+\dot u_{(\mu}u_{\nu)}\right] - \frac{1}{3} \theta \Delta_{\mu\nu}$, to the leading order in the perturbation $h_{ij}$, turns out to be
 \bea
 \theta&=&3H\,,\label{theta}\\
 \sigma_{ij} &=& \frac{1}{2} a^{2} \dot h_{ij}\,\label{sigma-ij},
 \eea
 respectively, where ${H}=\dot a/a$ and dot denotes derivative w.r.t cosmic time $t$ (see, \cite{Weinberg} for a detailed derivation). 
 
 Now by solving the Einstein's equations $G^{\mu \nu} = 8 \pi G T^{\mu \nu}$  to the linear order in $h_{ij}$ 
 we obtain the wave equation for the gravitational waves in a shear-viscous fluid. 
 Notice that the bulk viscosity only couples to scalar perturbations.
 The zeroth order equation for the $ij-$component of Einstein's equation $G_{ij}= 8 \pi G T_{ij} $ gives us
 \be
 - \frac{2\ddot a}{a} - H^{2} = 8\pi G (p-3\xi H)\,.
\label{zero}
 \ee
 where we have used eq.~(\ref{theta}).
 The first order equation $\delta G_{ij}= 8 \pi G \, \delta T_{ij} $ gives us
 \bea
&& \ddot h_{ij} + (3H+16\pi G \eta) \dot h_{ij}-\left[ \frac{4\ddot a}{a}+2H^{2}\right. \nonumber \\ 
&& \left. \hspace{1.6cm} +16\pi G (p-3\xi H) +\,\frac{\nabla^{2}}{a^{2}}\right] h_{ij} = 0\,,
%  \left[ h_{ij}^{\prime \prime} - \nabla^2 h_{ij} + 2 {\cal H} h_{ij}^\prime -2 h_{ij}( 2 {\cal H}^\prime+ {\cal H}^2 )  \right]a^{-2}  \nonumber\\
%  =  8 \pi G \left(2 p\,\, h_{ij} + 2 \eta \sigma_{ij} \right)
\label{first}
 \eea
where eq.~(\ref{sigma-ij}) has been used. Multiplying (\ref{zero}) by $2h_{ij}$ on both sides and then subtracting it from (\ref{first}), we obtain the wave equation for gravitational waves in a viscous fluid
\be  
\ddot h_{ij} + (3H+16\pi G \eta) \dot h_{ij}-\frac{\nabla^{2}}{a^{2}} h_{ij} = 0\,.
% h_{ij}^{\prime \prime} - \nabla^2 h_{ij} + 2 {\cal H} h_{ij}^\prime = - 16 \pi G a^2\, \eta\,\sigma_{ij} 
\label{hwave}
\ee 
% 
% The perturbation of the shear stress $ \sigma_{ij}$ is due the the perturbation of the fluid velocity by the gravitational waves. 
% The total four velocity is $\tilde u_\mu = u_\mu + \delta u_\mu$, normalizing the four velocity i.e., 
% $\tilde g^{\mu \mu} \tilde u_\mu \tilde u_\nu=-1$ in the the perturbed FRW metric 
% Eq (\ref{RW}), and retaining the first order terms in the metric and velocity perturbations gives us the relation 
% \be
% \delta u^\mu = \frac{1}{2} \delta g^{\mu \nu} \,u_\nu 
% \ee
% The spatial components of the velocity perturbations can be written in terms of the gravitational waves as 
% \be
% \delta v^i = h^{i j} v_{i}.
% \label{dv}
% \ee
% For the perturbed FRW metric the shear $\sigma_{\alpha \beta} \equiv (\tilde u_{(\alpha;\beta)}- (1/3) \theta \Delta_{\alpha \beta})$ and bulk expansion $ \theta \equiv{ { \tilde u}^\alpha}_{;\alpha}$ turn out to be
% \be
%  \sigma_{ij} =\frac{\partial}{\partial \tau} h_{i j}+ {\cal H} h_{ij}
% \label{sigmah}
% \ee
% and $\theta = 3 {\cal H}^\prime$ respectively to the leading order in the perturbation $h_{ij}$.

Going to the Fourier space and redefining the variable $h_{ij}$ as $\mu_{ij}/a$, the wave equation (\ref{hwave}) takes the form
\bea
 &&\ddot \mu_{ij} + (H+16\pi G \eta) \dot \mu_{ij}+ \bigg(\frac{k^{2}}{a^{2}}-\frac{\ddot a}{a}-H^{2} \nonumber \\
 &&  \hspace{3.5cm}   -16\pi G \eta H \Big) \mu_{ij} = 0.
\label{mu-wave}
 \eea

 In the conformal time $\tau$ defined through $dt=a\, d\tau$, the wave equation for $\mu_{ij}$ takes the form
 \bea
&&\mu_{ij}^{\prime \prime} + 16\pi G \eta \,a\, \mu_{ij}^{\prime} +\bigg(k^{2}-\frac{a^{\prime \prime}}{a}  \nonumber\\
&&  \hspace{2.8cm}-16\pi G \, \eta \,a\, {\cal H} \Big) \mu_{ij}= 0.
\label{mu-wave-conf}
 \eea
which on the subhorizon scales $k^{2}\gg \frac{a^{\prime \prime}}{a}$ reduces to 
 \bea
\mu_{ij}^{\prime \prime} + 16\pi G \,\eta \,a\, \mu_{ij}^{\prime} + k^{2} \mu_{ij} = 0.
\label{mu-wave-conf-reduced}
\eea
% \be
% h_{i j}^{\prime \prime} +{\bf q}^2 h_{ij} + 2 {\cal H} h_{ij}^\prime + 16 \pi G \eta\, a^2  \left( h_{i j}^\prime +{\cal H} h_{ij} \right) =0
% \ee
% Define $\mu_{ij}= a h_{ij}$. The wave equation for $\mu_{ij}$ is
% \be
% \mu_{ij}^{\prime \prime} + ({\bf q}^2 - \frac{a^{\prime \prime}}{a}) \mu_{ij}+ 16 \pi G\, a \eta\,\,  \, \mu_{i j}^\prime =0
%  \ee
%  
%  For gravitational waves with wavelengths smaller than the cosmological horizon $ |{\bf q}|\,\tau \gg 1$, the $a^{\prime \prime}/a$ term is small compared to $q^2$ and the wave equation for $\mu_{ik}$ reduces to the form
%  \be
% {\ddot \mu_{ij} }+ {\bf k}^2 \mu_{ij} + 16 \pi G  \eta\,\,  \,{\dot \mu_{i j}}=0
%  \ee
% where overdot represents derivative w.r.t cosmic time $dt = a d\tau$ and $|{\bf k}| = |{\bf q}|/a$ is the physical wave number of the gravitational waves.

The amplitude of the radial component of the wave $A_{\times,+}= r \mu_{ij}$ of the two polarization modes $\times$ and $+$ satisfies the following one dimensional wave equation at large distances from the source
\be
\ddot{A} + \beta \, a\, \dot{A} + k^2 A = 0 \; \label{amp},
\ee
where $\beta\equiv16\pi G \, \eta$. Let the solution of (\ref{amp}) be
\be
A(\tau,\omega) = \tilde A(\omega)\,e^{i k r - \int i \omega d\tau} \;,
\ee
and substituting it in the equation (\ref{amp}) gives us the dispersion relation
% On taking a temporal Fourier transform 
% \be
% A(t,x) = \int_{-\infty}^{\infty} \frac{d \omega}{2 \pi} ~e^{i (kx - \omega t)} ~\tilde{A}(\omega) \; ,
% \ee
% and substituting in the above differential equation, we find that 
 \be
 - \omega^2 - i \beta \,a\,  \omega + k^2 = 0\; .
 \ee
Writing $k$ in terms of real and imaginary parts $k = k_R + i k_I$, the above dispersion relation gives us (using the weak damping approximation $\beta \ll \omega$, and retaining only leading order terms)
 \begin{eqnarray}
 k_R = \omega \;, ~~~ k_I = \frac{\beta \, a}{2} \; .  
 \end{eqnarray}
The real part of $k$ is $\omega$ so that there is no dispersion at this order but the presence of imaginary part of $k$ 
causes attenuation of the wave. Substituting for $k = k_R + i k_I$ into eq.~(\ref{amp}), the solution becomes
\be
A(\tau,\omega) = \tilde A(\omega)\,e^{i k_{R} r - \int i \omega d\tau} \times e^{-k_I r}\;,
\ee
therefore the strain $h_{ij}$ measured at the detector $L=a\,r$, in the cosmic time $t$, will be 
 \be \label{eq:detector}
h_{ij} = \frac{\tilde A(\omega(t))}{L_0}\,e^{i k_{R} r - \int i \omega_{p} dt} \times \frac{L_0 \, e^{-\frac{\beta}{2} L}}{L} \; .
\ee
where $\omega_{p}=\frac{\omega}{a}$ is the physical angular frequency and $L_0$ represents the source distance for zero shear viscosity. 
The attenuation of the GW due to shear viscosity after traveling over a distance of $L=a\,r$ is by the factor $L_0\, e^{-\frac{\beta}{2} L}/L$.
The attenuation length is $k_I^{-1} =  \eta^{-1} M_{\rm Pl}^2$. Since the strain measured depends upon the
masses of the binary black holes through the combination called chirp mass $M_c = \frac{(m_1 m_2)^{3/5}}{(m_1 + m_2)^{1/5}}$
% \be
% M_c = \frac{(m_1 m_2)^{3/5}}{(m_1 + m_2)^{1/5}}
% \ee
which is determined from the observed GW frequency as a function of time \cite{Husa:2009zz}
\be
\dot f_{gw} = \frac{96 \pi^{8/3}}{5} \left(\frac{G M_c}{c^{3}}\right)^{5/3}\,  f_{gw}^{11/3}\,.
\ee
We note that it is expected that the effect of viscosity will be degenerate with the effect of a number of other parameters such as source distance, orientation of the plane of the binary BH system etc. In this work, we aim to illustrate how constraints on viscosity could be obtained, thus, we have fixed the values of these parameters and studied only the constraints on viscosity and source distance. It is expected that in future, with the help of the upcoming virgo detector (and other GW observatories), improved triangulation methods could help constrain the source distance a lot better and this could be used to improve the limits on shear viscosity.
We fit the two remaining variables $L$ and $\eta$ in our model to the strain data as measured in \cite{LIGO}.

\begin{figure}
  \includegraphics[width = .45\textwidth]{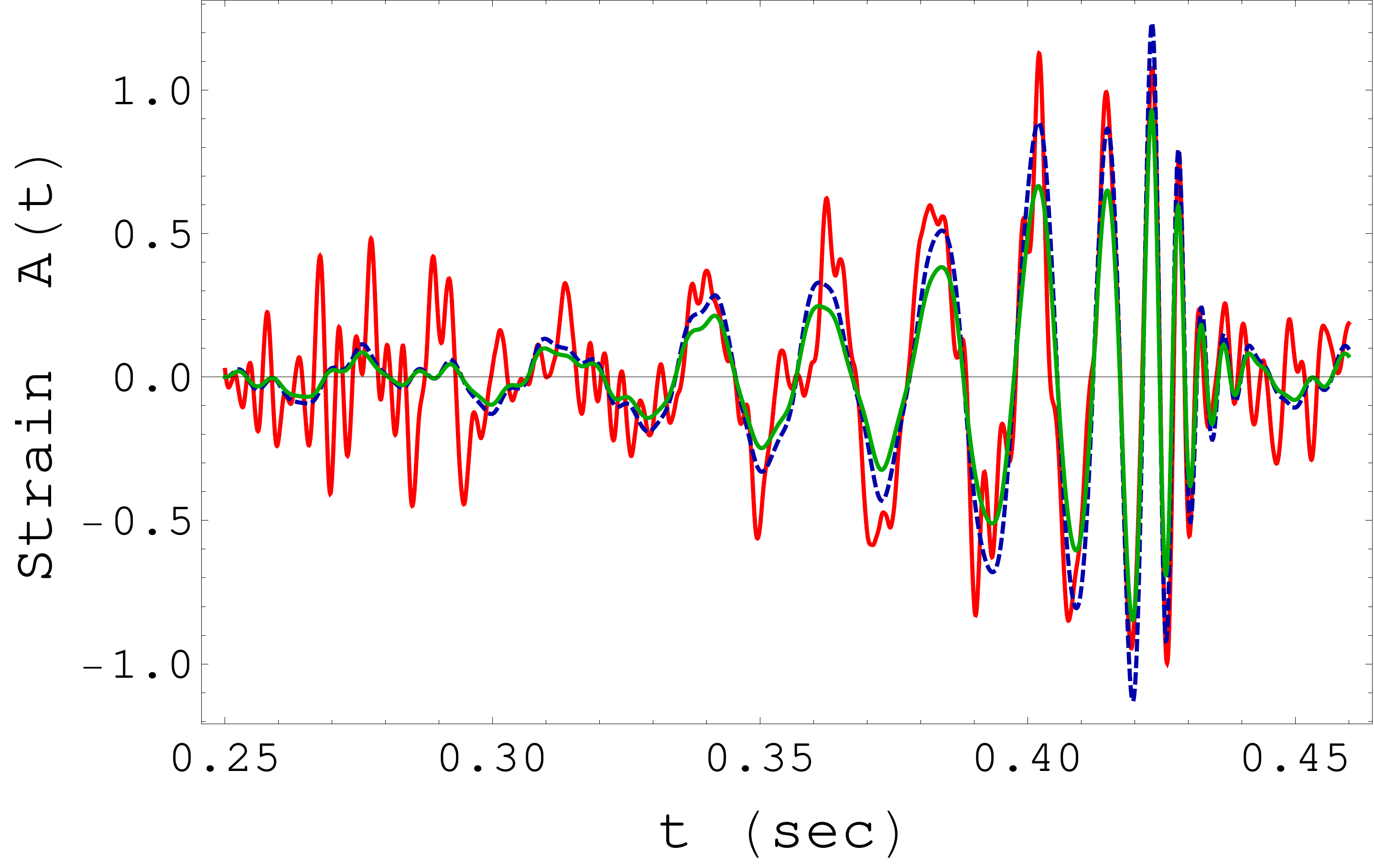}
  \caption
  {The effect of viscosity on the GW strain time series:
  The red curve is the time series of strain of the observed GW data and the blue (dashed) curve is the theoretical strain when $Q=0$.
  When $Q = 1$, green curve, the theoretical strain gets attenuated.
%   {\color{green} and its corresponding likelihood will be lesser.
%   Notice that for the signal corresponding to 
%   the inspiral phase and the ringdown phase, the effect of viscosity of intervening medium is little but for the merger phase, the effect of viscosity is much more pronounced.} 
  All the time series are band-limited to the frequency range $30-350 \, Hz$ and the data is from the LIGO Hanford detector. 
  }
  \label{time-series}
\end{figure}

For dark matter and dark energy, the magnitude of the shear viscosity stress can be given by
\be
T^{ij}_{\rm viscous} \simeq \frac{\eta_{\rm crit}}{H_0^{-1}} \simeq \rho_{\rm crit} \; ,
\ee
which defines $\eta_{\rm crit} = \rho_{\rm crit} H_0^{-1}= 3.21 \times 10^{-5} ({\rm GeV})^3 = 4.38 \times 10^{8} {\rm Pa}\,\, {\rm sec}$. 
Writing the actual $\eta = Q \eta_{\rm crit}$, we can put constraints on the dimensionless number $Q$ 
using GW observations.
Using $\rho_{\rm crit} = 3 H_0^2 M_{\rm Pl}^2$,
the attenuation of GW after traveling over a distance $L$ is given by the factor 
\be
\frac{L_0\, e^{-\frac{\beta}{2} L}}{L} = \frac{L_0 \, e^{ -3 Q L H_0 }}{L}
\ee
Since $L \sim {\cal O} (0.1) H_0^{-1}$ which implies $3 Q L H_{0}\sim{\cal O} (0.1)\times 3Q$. Thus, assuming $L=L_0$ and the viscosity of the cosmic fluid $Q \sim {\cal O} (1)$ amounts to, $\frac{L_0 e^{-3 Q L H_0}}{L} \sim 0.75$, $25\%$ attenuation of GW amplitude, see fig.(\ref{time-series}).
% Thus, the viscosity of the cosmic fluid of the order one in units of $\rho_{\rm crit} {H_0}^{-1}$ in order to have an observable effect on GWs 
% which changes the amplitude of the wave by a factor $e^{-k_I L} = e^{-{\cal O} (0.1)} \approx 0.9$ i.e. ten percent attenuation. 
% However due to instrument noise the bound on $\eta$ we will derive from the analysis of GW150914 turns out to be less stringent than 
% $\eta <  {\cal O}(0.1) \eta_{\rm crit}$.
% For $k_I L \sim {\cal O} (0.1)$, and since $L \sim {\cal O} (0.1) H_0^{-1}$, we have $Q \sim {\cal O} (1)  \; .$
% Thus, the viscosity of the cosmic fluid must be of the order one in units of $\rho_{\rm crit} {H_0}^{-1}$ in order to have an observable effect on GWs 
% which changes the amplitude of the wave by a factor $e^{-k_I L} = e^{-{\cal O} (0.1)} \approx 0.9$ i.e. ten percent attenuation. However due to instrument noise the bound on $\eta$ we will derive from the analysis of GW150914 turns out to be less stringent than 
% $\eta <  {\cal O}(0.1) \eta_{\rm crit}$.
We will now explain how one could use the data publicly released by the LIGO collaboration and use the attenuation factor obtained in the above equation to constrain the cosmic viscosity.

\vspace{0.3cm}

{\it Data analysis:} 
Eq (\ref{eq:detector}) implies that if the intervening fluid has a non-zero viscosity, the amplitude of observed GWs should be lower 
(see fig (\ref{time-series}) ) so that we can use the strain observations to put limits on $Q$.
In order to proceed, we need to find the change in a convenient measure of goodness of fit as we change the viscosity of the cosmic fluid.
The observed strain will be different from the theoretically predicted strain due to the presence of noise 
\be
A^{\rm obs}(t) = A^{\rm th}(t) + n(t) \; .
\ee

We use the data obtained by the LIGO Hanford detector for the gravitational-wave event GW150914 on September 14, 2015 at 09:50:45 UTC
provided by The LIGO Open Science Center \cite{LOSC}.
The released data provides the strain observations for a time interval $T = 0.21$ secs and within this time, 
it has been sampled 3340 times which implies a sampling rate 
of $16384$ per sec and the corresponding Nyquist critical frequency of 8192 Hz.  
The sub-interval size in the  frequency domain is 4.7628 Hz, which is also the minimum frequency, the maximum frequency being the Nyquist frequency.
All the released  time series data has been filtered with a $35-350$ Hz bandpass filter to suppress large fluctuations outside the 
detector's most sensitive frequency band, we thus restrict ourselves to this range of frequencies for the rest of the analysis. 

%At times $t_{j}$ and $t_{j'}$, the noise values are given by two random variables $n(t_{j})$ and $n(t_{j'})$ which are, in general correlated. 
%If we consider $N$ such time values, then we have $N$ random variables for the noise values which are not independent so that 
%there will be a joint probability distribution function in this $N$ dimensional space, which, for the case of stationary noise will be invariant 
%under translations in time. 
If the joint distribution of the noise values at different values of time is a Gaussian, then, given a set of theoretical 
strains $A(t_{j})$ the Likelihood function (the probability of data, given the theory) will be given by
\be \label{eq:gaussian-likelihood}
{\cal L} = \frac{1}{( (2\pi)^N \det C_{j j'})^{1/2}} \exp \bigg\{ - \frac{1}{2} ~\sum_{j j'}~  \xi_j ~ C_{j j'}^{-1} ~ \xi_{j'} \bigg\} \; ,
\ee
where
$\xi_j = A^{\rm th}(t_j) - A^{\rm obs}(t_j)$ is the difference between theoretical signal and the observed signal while 
$C_{jj'}$ is the noise covariance matrix. 
The noise could in general be non-stationary and non-Gaussian due to the presence of glitches (i.e. noise transients, see e.g. \cite{Non-Gaussian})
but here we proceed assuming Gaussianity.
%assume that the such glitches have already been removed from sensitivity data provided.
For stationary noise, the noise covariance matrix will be diagonal when we transform to the frequency domain:
\be \label{eq:likeli}
\langle {\tilde n}(f) {\tilde n}^*(f') \rangle = \frac{1}{2} \delta(f-f') S_n(f) \; ,
\ee
where $S_n(f)$ is the Power Spectral Density (PSD) of the noise background of the detector.
Thus, in frequency domain, the Likelihood function will be given by an expression similar to Eq(\ref{eq:likeli})
except for the fact that the matrix $C$ will be diagonal and can be readily inverted.

The LIGO Open Science Center \cite{LOSC} has also released the average measured strain-equivalent noise, or sensitivity, of the Advanced LIGO detectors during the time analyzed (i.e. Sept 12 - Oct 20, 2015). The frequency range 0 to 8192 Hz (the Nyquist frequency) has been divided into 65536 sub-intervals each of size $\Delta f = 0.125$ Hz and the Amplitude Spectral Density (ASD) i.e. $\sqrt{ S_n(f)}$ is provided for each of the  intervals. From ASD, we can readily obtain $S_n(f)$, the PSD.
The step size in frequency domain for observed strain and the theoretical strain is 4.7628 Hz while the step size in frequency domain for ASD is 
0.125 Hz. We integrated the PSD 
\be \label{eq:noisesd}
\int_{f_1}^{f_1+\Delta f_1} S_n(f) ~df = \sigma_n^2 (f_1)\; ,
\ee
to obtain the noise variance at each of the frequencies at which we have the theoretical and observed strains. These are the non-zero elements of the 
noise covariance matrix which is diagonal in frequency domain. 
The typical values of the measured strain are of the order of $10^{-21}$ while typical values of $\sqrt{S_n(f)}$ in the 
frequency range 20-450 Hz are of the order of $10^{-23}$. 
%Fig (\ref{fig-sigma_diff}) shows the corresponding values of 
%$\log |\tilde{A}^{\rm th}(f_k) - \tilde{A}^{\rm obs}(f_k)|$ and $\log \sigma_n (f_k)$ for all the 92 frequency values in the range 20-450 Hz.
%One can now readily evaluate the exponential in Eq(\ref{eq:gaussian-likelihood}) and hence evaluate a relative likelihood.
%\begin{figure}[htbp]
%  \includegraphics[width = .45\textwidth]{sigma-diff}
%  \caption
% {Solid blue is $\log |\tilde{A}^{\rm th}(f) - \tilde{A}^{\rm obs}(f)|$ while dashed green is $\log \sigma_n (f)$ at the data points in the frequency range
% 20-450 Hz using the data from LIGO Hanford detector. All the quantities are expressed in units of $10^{-21}$. }
%  \label{fig-sigma_diff}
%\end{figure}

\begin{figure}
  \includegraphics[width = .45\textwidth]{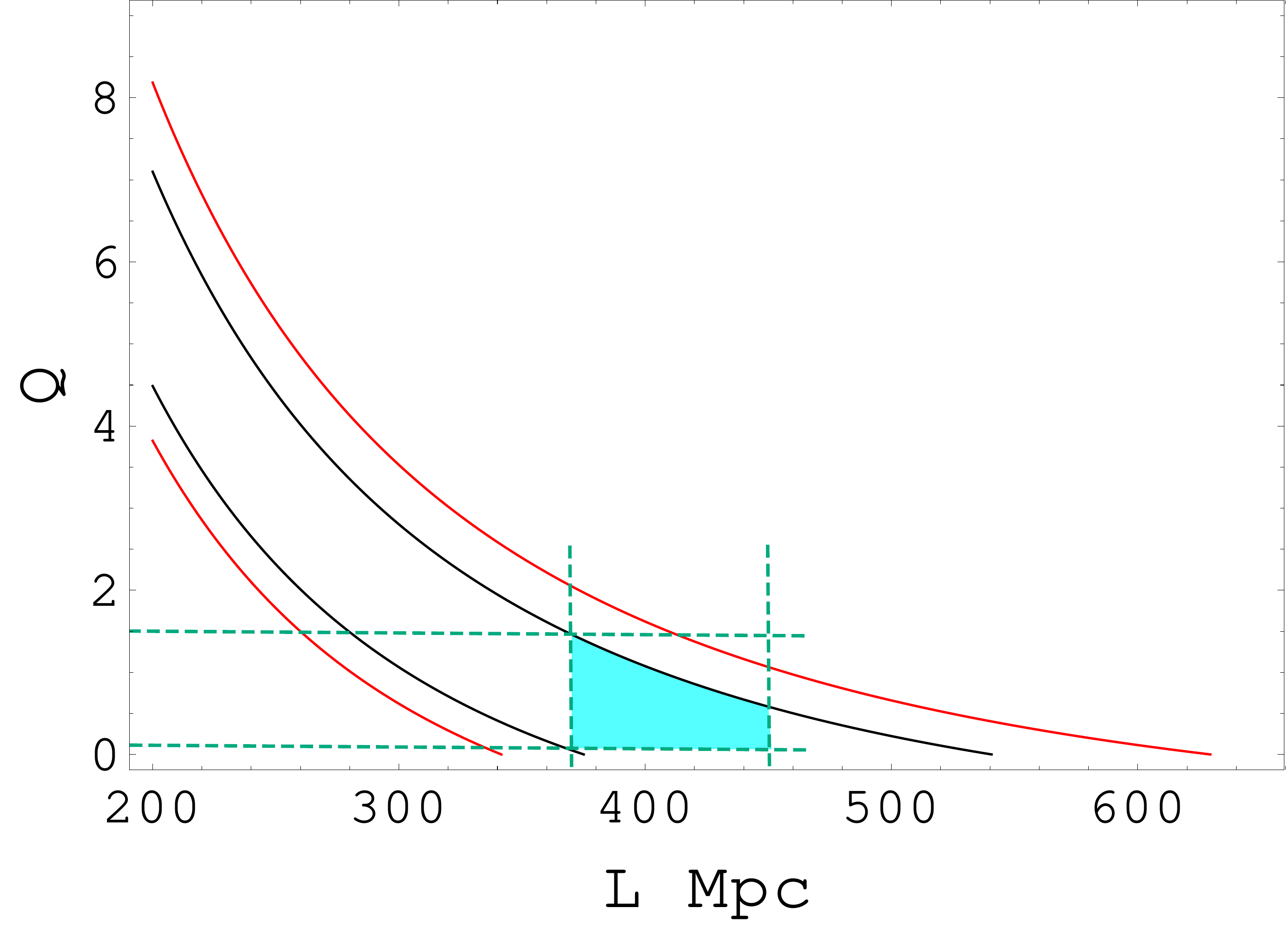}\vspace{0.5 cm}
  \includegraphics[width = .45\textwidth]{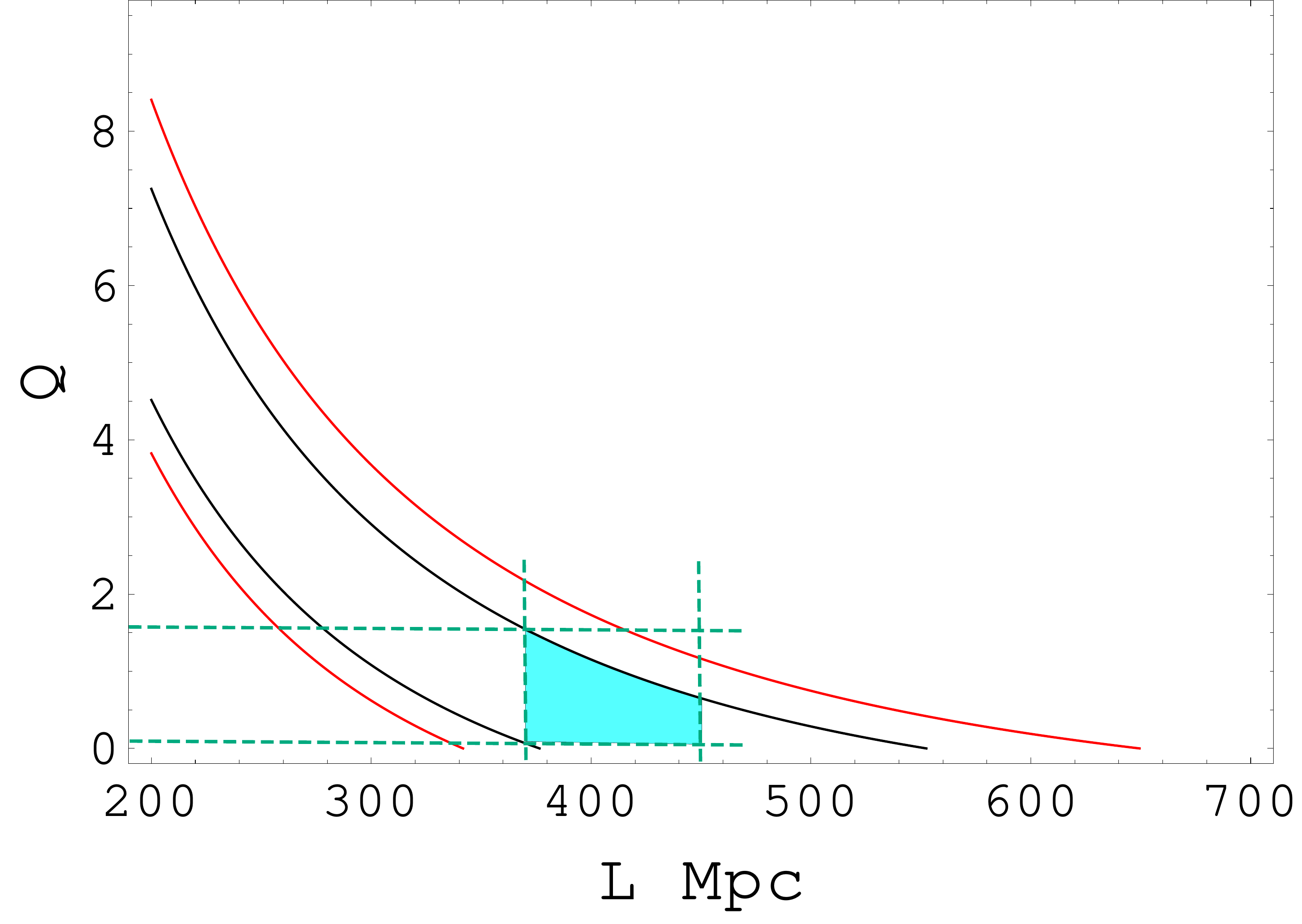}
%  \captionsetup{justification=raggedright,singlelinecheck=false}
  \caption
 {The constraints on $Q$ and $L$ at $1\,\sigma \,\, CL$ (inner, black contour) and $2\, \sigma \,\, CL$  (outer, red contour) for the data obtained from Hanford (upper panel) and Livingston (lower panel) observatories. 
The shaded region between the two vertical lines is the source distance with an uncertainty of 10\% around the estimated central value of $L=410$ Mpc, the corresponding range of values of $Q$ can be easily seen.
}
  \label{fig_likelihood_freq}
\end{figure}

However, it turns out that the most important source of error is not the detector noise, but the statistical error due to a finite time of observation. Working in frequency domain, for the $92$ frequency values of interest, the root mean square fluctuation in the observed signal can be estimated by the sample mean of $\sqrt{|A^{\rm obs} (f)|^2}$. Using this as the expected number of events in the definition of $\chi^2$ statistic, one can find the $\chi^2$ per degree of freedom due to the statistical error. Fig.(\ref{fig_likelihood_freq}) shows the contour plots of constant value of this $\chi^2$ per degree of freedom as the parameters $Q$ and the distance to the source $L$ are varied. 
%{\color{blue} 
The black and red contours are the boundaries of regions within 1-$\sigma$ and 2-$\sigma $, respectively, of the parameter values which minimize the $\chi^2$. Note that for $Q=0$ and $L=410$ Mpc, the value of $\chi^2$ is 22.15 for the data obtained Hanford observatory and 28.41 for the data obtained from Livingston observatory. 
In fig (\ref{fig_likelihood_freq}), the contours corresponding to $1\,\sigma \,\, CL$ (inner, black contour) are arrived at by finding the combination of $L$ and $Q$ which increases the $\chi^2$ by 2.3 while those at $2\, \sigma \,\, CL$  (outer, red contour) are arrived at by finding the combination of $L$ and $Q$ which increases the $\chi^2$ by 6.18.
%}
%{\color{red}(recall that 1.6 sigma standard deviation corresponds to 90 \% bayesian confidence level).}

For every choice of source distance, we can find a value of viscosity. Thus, an independent knowledge of the distance of the source could help in determining the limits on viscosity better. From fig.(\ref{fig_likelihood_freq}), it is clear that the source distance estimated by the LIGO collaboration corresponds to a nearly vanishing value of viscosity but one can find the upper bound on the distance to the source. It can be inferred from the fig.(\ref{fig_likelihood_freq}) that shear viscosity of the cosmological fluid in the path of GW150914 has the upper bound $\eta \lesssim 5.2 ~\eta_{\rm crit} ~\approx~ 2.3 \times 10^{9} ~{\rm Pa }\,\,{\rm sec}~ {\rm at}~ 1~ \sigma~ CL$, if the luminosity distance of the source is fixed to the value $L=(410-180)\,Mpc=230\, Mpc$ which is the lower limit on the distance of the source by the observation $L=410^{+160}_{-180}\, Mpc$ ~\cite{LIGO}. If the GW events in future can be located by independent observations of their electromagnetic signals and the distance fixed to say $10\%$ accuracy then, as shown in fig.(\ref{fig_likelihood_freq}), the value of $Q$ can be much more restricted compared to the present constraints.

% For this value of distance, notice that the trough of the $\chi^2$ function at $Q \neq 0$ is due to statistical fluctuation 
% and has a statistical significance of less than $(1/10) \sigma$. Similar limits can be found for any value of distance. 

% \begin{figure}
%   \includegraphics[width = .45\textwidth]{likelihood_freq-domain-result-2}
% %  \captionsetup{justification=raggedright,singlelinecheck=false}
%   \caption
%  {The change in likelihood as one changes $Q$, the dimensionless cosmic viscosity for the various values of the source distance.
%  For the central value of distance (410 Mpc) of the source, the solid (red) curve is obtained by using the data from 
%  the Hanford detector while the dashed (blue) curve is for the Livingston detector.
%  For the lower value of distance (230 Mpc) of the source, the dotted (brown) curve is obtained by using the data from 
%  the Hanford detector while the dot-dashed (purple) curve is for the Livingston detector.
%  }
%   \label{fig_likelihood_freq}
% \end{figure}

\vspace{0.3cm}

{\it Self interacting dark matter in galaxies and  clusters:} 
Self interaction of dark matter has been introduced in models of dark matter \cite{DM1,DM2,Fan:2013yva,Foot:2014uba,Foot:2016wvj} for solving the Core-Cusp problem of galaxies  \cite{Core-Cusp}, the problem of galactic substructure \cite{Substructure} and/or the Missing-Satellites problem \cite{Missing-Satellites}.  
The self interaction cross section can be related to the shear viscosity of DM 
by the relation $\eta= (1/3) m\, n\, v\, l$ where  the mean free path of DM particles $l$ can be related to its number density  $n$ and self interaction cross section $\sigma$ as $l=1/ (n \sigma)$ and the shear viscosity of DM is $\eta= (1/3) (v\,m/\sigma)$.
For self-interacting dark matter in galaxies the mean free path $l\sim 100\, kpc$, $\rho= m n \sim 0.4\, {\rm  Gev/cm^3}$ and $v=220\, km/sec$ and the typical value of the shear viscosity of dissipative dark matter is $\eta \sim 10^7 \,{\rm Pa\,\,sec}$.

In a recent study of the galaxy cluster Abell 3827 \cite{Massey:2015dkw}, four elliptical galaxies are observed to fall towards the center of the cluster and there is an offset between the dark matter (inferred from lensing) and the visible matter which can be ascribed to a self interaction between dark matter, the corresponding cross section by mass value is estimated to be
$\sigma/m = (1.7 \pm 1) \times 10^{-4} {\rm cm}^2/{\rm gm}$. 
 Using the $m/\sigma$ value inferred from Abell 3827, the shear viscosity of DM at cluster scales has the value $\eta= 5.9 \times 10^{9}~ {\rm Pa} \,\,{\rm sec}$. This value of $\eta$ is close to the constraint  $\eta \lesssim 2.3\times 10^{9} ~ {\rm Pa}\,\, {\rm sec}$ inferred from the analysis of GW150914. A more refined estimate \cite{Kahlhoefer:2015vua}  of the DM self interaction in Abell 3827  (which takes into account that the DM is gravitationally bound in the cluster) gives the cross section by mass as 
$\sigma/m= 1.5 ~{\rm cm}^2/{\rm gm}$. This results in a lower value of shear viscosity $\eta= 0.6 \times10^{6}~ {\rm Pa}\,\,{\rm sec}$. 

\vspace{0.3cm}

{\it Conclusions:}
In this work, we explored the possibility that the upcoming observations of gravitational waves could observationally constrain the viscosity of cosmic fluid. We began by deriving the effect of cosmic shear viscosity on the propagation of Gravitational Waves in the Universe. 
Except for the source distance and viscosity, we fixed the values of all the other parameters of the binary black hole system observed by the LIGO collaboration. We found that if the distance to the source can be independently determined, one can, at least in-principle, put interesting upper limits on the shear viscosity of the medium intervening the source and the point of observation.
Our results are best interpreted as a proof-of-principle demonstration of how this could be done.
Thus, we put constraints on the shear viscosity of dark matter and dark energy which makes such models testable. 
Future observations of GW at LIGO, VIRGO, LISA and other observatories could potentially probe the viscous properties of cosmological fluids and will be able to verify or rule out these models of cosmology. 

 \noindent{\bf Acknowledgements}
GG would like to thank Anand Sengupta (IIT-Gandhinagar), Mudit Srivastava (PRL, Ahmedabad) and 
Jayanti Prasad (Savitribai Phule Pune University, Pune) for discussions.
 \\ \\

\end{document}